%% file: apssamp.tex
\documentclass[%
 reprint,
 amsmath,amssymb,
 aps,
prd
]{revtex4-2}

\usepackage{graphicx}
\usepackage{dcolumn}
\usepackage{bm}

\begin{document}

\preprint{APS/123-QED}

\title{LearningMatch: Siamese Neural Network Learns the Match Manifold}

\author{Susanna Green}
\affiliation{Institute of Cosmology and Gravitation, University of Portsmouth, Portsmouth PO1 3FX, United Kingdom}

\author{Andrew Lundgren}
\affiliation{Catalan Institution for Research and Advanced Studies (ICREA), E-08010 Barcelona, Spain}
\affiliation{Institut de F´ısica d’Altes Energies (IFAE), The Barcelona Institute
of Science and Technology, UAB Campus, E-08193 Barcelona, Spain}
\affiliation{Institute of Cosmology and Gravitation, University of Portsmouth, Portsmouth PO1 3FX, United Kingdom}

\author{Xan Morice-Atkinson}
\affiliation{Institute of Cosmology and Gravitation, University of Portsmouth, Portsmouth PO1 3FX, United Kingdom}

\date{\today}


\begin{abstract}
\input{abstract}
\end{abstract}

\maketitle

\section{Introduction}
\input{Introduction}

\section{The Match}
\input{Theory}

\label{Theory}

\section{LearningMatch}
\input{LearningMatch}
\label{LearningMatch}

\section{Results}
\input{Results}
\label{Results}

\section{Discussion}
\input{Discussion}
\label{Discussion}

\section{Conclusions}
\input{Conclusions}

\label{Conclusions}


\begin{acknowledgments}
Thank you to Laura Nuttall and Ian Harry for their useful comments. Susanna Green was supported by a STFC studentship and the University of Portsmouth. Andrew Lundgren acknowledges the support of UKRI through grants ST/V005715/1 and ST/Y004280/1. Xan Morice-Atkinson acknowledges the support of UKRI through grants ST/V005715/1 and ST/Y004280/1. Numerical computations were done on the Sciama High Performance Compute (HPC) cluster which is supported by the ICG, SEPNet and the University of Portsmouth. This paper has been assigned document number LIGO-P2500007.
\end{acknowledgments}

\bibliography{references}

\end{document}

%% file: abstract.tex
The match, which is defined as the the similarity between two waveform templates, is a fundamental calculation in computationally expensive gravitational-wave data-analysis pipelines, such as template bank generation. In this paper we introduce LearningMatch, a Siamese neural network that has learned the mapping between the parameters, specifically $\lambda_{0}$ (which is proportional to the chirp mass), $\eta$ (symmetric mass ratio), and equal aligned spin ($\chi_{1}$ = $\chi_{2}$), of two gravitational-wave templates and the match. The trained Siamese neural network, called LearningMatch, can predict the match to within $3.3\%$ of the actual match value. For match values greater than 0.95, a trained LearningMatch model can predict the match to within $1\%$ of the actual match value. LearningMatch can predict the match in 20 $\mu$s (mean maximum value) with Graphical Processing Units (GPUs). LearningMatch is 3 orders of magnitudes faster at determining the match than current standard mathematical calculations that involve the template being generated. 

%% file: Introduction.tex
 Since the first neural network architecture in the 1970's, many neural network models have been designed to tackle various problems and data sets such as LeNet-5, AlexNet, and ChatGPT \citep{Fukushima1980, FukushimaandWake1991, LeCun1998, Krizhevsky2012, ChatGPT}. Neural networks have a unique property in which they are able to approximate any continuous function in 2 layers which is called the \textit{universal approximation} theorem \citep{cybenko1989}. This property and the fact that neural networks can now be trained in a reasonable amount of time due to the back propagation algorithm and readily available Graphical Processing Units (GPU), has led to neural networks being utilised in a variety of fields including gravitational-wave data analysis \citep{Rumelhart1986LearningErrors}. 
 
The use of machine learning algorithms in gravitational-wave astronomy was originally suggested in 2006 \citep{Lightman2006}. In recent years, neural networks have been widely applied to a variety of problems in the gravitational-wave community: from fast parameter estimation \citep{Andres-Carcasona2024FastNetworks, Dax2021Real-TimeEstimation, McLeod2022RapidLearning, Gabbard2022BayesianAstronomy, Green2021, Chua2020, Daniel2018_realdata} to solving the Teukolsky equation \citep{Luna2023SolvingNetworks}, from categorising noise \citep{Alvarez-Lopez2024GSpyNetTree:Candidates, Fernandes2023ConvolutionalStreams, Razzano2018Image-basedDetectors, Bahaadini2017DeepClassification, Zevin2017} to waveform modelling \citep{Luna2024NumericalStars, Thomas2022AcceleratingNetworks, Freitas2022GeneratingLearning}, from searching for exotic gravitational wave signals \citep{Verma2022DetectionNetwork, Attadio2024AWaves, Ravichandran2023RapidLearning, Qiu2023DeepMergers, Iess2023LSTMData, Morras2022SearchDetectors, Iess2020Core-CollapseClassification, Dreissigacker2020} to confirming evidence of precession in current events \citep{Macas2024RevisitingMitigation}. In this research we introduce a type of neural network, called a Siamese neural network, to the gravitational-wave community \cite{J2020ANetworks}. Siamese neural networks were originally used to verify signatures and since then have been applied to a variety of problems that require a similarity metric to be learned, for example face verification \cite{BROMLEY1993SIGNATURENETWORK, Chopra2005LearningVerification}. This neural network architecture has been designed to learn the `similarity' in data and for this reason we have used it to learn the similarity between two gravitational-wave templates, otherwise known as the match \citep{BenOwen1996}. Similar research has been conducted using neural networks to predict the mismatch for the optimisation of numerical relativity simulations placement \citep{Ferguson2023OptimizingNetwork}. The aim of this research is to design a Siamese neural network that could be integrated into gravitational-wave data analysis pipelines, because many of these algorithms are computationally expensive with regards to time taken and computational hardware \citep{Stergioulas2024MachineAstronomy}. 

Over 90 gravitational-wave events have been observed since the first binary black hole merger in 2015 by LIGO, Virgo, and KAGRA \citep{Abbott2016GW150914:Discoveries, Abbott2019GWTC-1:Runs, Abbott2021GWTC-2:Run, Abbott2023GWTC-3:Run}. These gravitational waves have been from a variety of astrophysical sources including an asymmetric binary black hole merger, an intermediate-mass black hole inspiral, and neutron star-black hole candidates \citep{Abbott2020GW190412:Masses, Abbott2020GW190521:M_odot, Abbott2020GW190814:Object}. Many of these gravitational-wave events have been extracted from the interferometric strain data using template banks \citep{Allen2021OptimalBanks, Adams2016, PyCBCSearch2016, Kovalam2022EarlySearch, Sakon2024TemplateKAGRA}. Template banks are a catalogue of waveform models (i.e. templates) with different parameters that approximate the actual gravitational wave signal \citep{Allen2021OptimalBanks, Ajith2007AWaveforms}. Template banks are computationally expensive to generate because the match is constantly being calculated in many template bank generation algorithms. For future gravitational-wave detectors, template bank generation is going to be unfeasible because the templates are of the order of hours or days in length and therefore the match calculation is going to be a computationally intensive \citep{Lenon2021EccentricExplorer}. 

In this paper, we introduce a Siamese neural network called LearningMatch, that has learned the match between two templates and therefore has the potential to be integrated into template bank generation algorithms. This paper is organised as follows. In Section \ref{Theory} the match is explained while in Section \ref{LearningMatch} we describe a Siamese neural network that is required to predict the \textit{match}. The accuracy of LearningMatch is showcased in Section \ref{Results} with potential applications of LearningMatch highlighted in Section \ref{Discussion}. This research will then be summarised in section \ref{Conclusions}. All the code will be open source and accessed at \href{https://github.com/SusannaGreen/LearningMatch}{https://github.com/SusannaGreen/LearningMatch}. The datasets for this research can be accessed on Zenodo \citep{susanna_green_zenodo}.

%% file: Theory.tex
\begin{figure}
    \centering
    \includegraphics[scale=0.4]{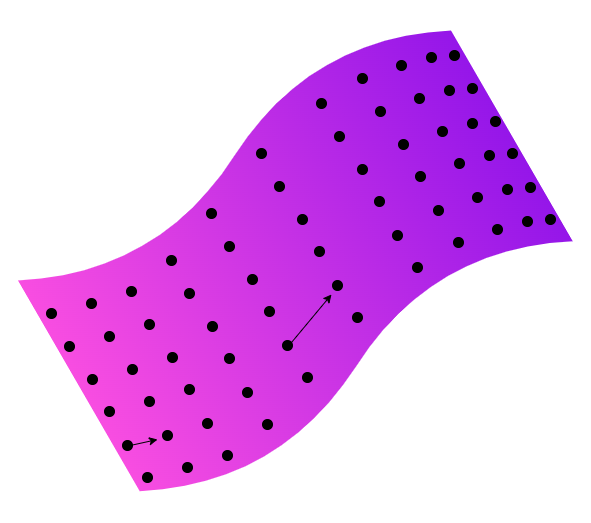}
    \caption{A schematic diagram of the match manifold between two gravitational wave templates. The black dots represent an equal change in a parameter, for example mass, of a binary black hole merger. The distance between these black dots represents the match. The arrows highlight the fact that when a parameter is changed equally, the distance varies depending where it is on the match manifold. In binary black hole mergers when the mass is changed by the same amount, the match changes drastically in the low mass regions compared to the high mass regions.}
    \label{fig: Match Manifold}
\end{figure}

The match is defined as the weighted inner product between two gravitational-wave templates \citep{BenOwen1996}. Gravitational wave signals emitted by mergers of compact objects, such as binary black holes, are modelled by templates. Templates are generated by dividing the gravitational-wave signal into three phases (specifically inspiral, merger, and the ringdown) because the physics used to explain a coalescing compact binary is different at each stage. In the inspiral phase, a stationary phase approximation is used and the gravitational signal, \textit{h(f)}, is described by 

\begin{equation}
h(f) = A(f) e^{i \Psi(f)}, 
\label{gravitational-wave}
\end{equation}

where $\Psi(f)$ is the phase of the coalescing compact binary and \textit{A(f)} is defined as,

\begin{equation}
A(f) = \sqrt{\frac{5}{24}} \frac{\mathcal{M}_{c}^{-5/3}}{\pi^{2/3} D} f^{-7/6},
\end{equation}

to leading order (\textit{D} is the distance, $\mathcal{M}$ is the chirp mass (see equation \ref{chirp mass}) and \textit{f} is the frequency) \citep{Buonanno2009ComparisonDetectors}. In Equation \ref{gravitational-wave}, the phase, $\Psi(f)$, is defined as as 

\begin{align*}
\Psi(f) = 2\pi t_{c}f - \phi_{c} - \pi/4
+  \sum_{j=0}^{7} \psi_{\frac{j}{2}PN} f^{(-5+j)/3} \\
+ \sum_{j=5}^{6} \psi^{l}_{\frac{j}{2}PN} ln(f) f^{(-5+j)/3},
\label{phase expansion}
\end{align*} 

to 3.5 Post-Netwonian order. The Post-Newtonian (PN) approximation is used to model the coalescing compact binaries because it is assumed they are in a circular orbit, point masses and orbiting slowly so a velocity expansion can be used \citep{Bernard2009ARelativity, PoissonandWill1995}. The two terms, $\psi_{\frac{j}{2}PN}$ and $\psi^{l}_{\frac{j}{2}PN}$ are defined as,

\begin{equation}
\label{GW lambda}
\psi_{0} = \frac{3}{128}\pi^{5/3}\mathcal{M}_{c}^{5/3},
\end{equation}

to first-order expansion (see reference for more detail) \citep{Buonanno2009ComparisonDetectors}. So far, the inspiral phase of a compact binary merger has been discussed. The second phase, the merger, is modelled using numerical integration or interpolating between the inspiral and ringdown. The ringdown is modelled as a rotating black hole, see references for more information \citep{Teukolsky1973PerturbationsPerturbations, Teukolsky1974}. All three phases are then combined to form the gravitational-wave templates, such as IMRPhenomXAS, used by the gravitational-wave community \citep{IMRPhenomXAS}.

Templates that model compact binary coalescence are modelled by 15 parameters (where eccentricity is negligible) which can be decomposed in two categories: extrinsic and intrinsic parameters \citep{CurtandFlanagan1994}. The intrinsic parameters, $\theta$ (which includes the mass and spin of the two bodies), impact the actual shape of the waveform. The extrinsic parameters, $\phi_{c}$ and  $t_{c}$ (phase and time of coalescence respectively), affect the amplitude and phase of the waveform. The \textit{ambiguity function} measures how distinguishable two waveforms are, $h(\lambda)$ and $s(\Lambda)$, in the absence of any noise and is expressed as the inner product 
\begin{equation}
    A(\lambda, \Lambda) = (h(\lambda)|s(\Lambda)), 
\end{equation}
where $\lambda$ and $\Lambda$ are the parameter vectors of the template and signal respectively \citep{Sathyaprakash1999}.

$\lambda = (\phi_{c}, t_{c}, \theta)$ can be further divided into intrinsic parameters, $\theta$, and extrinsic parameters, $\phi_{c}, t_{c}$. In the context of searching for gravitational wave signals, the \textit{ambiguity function} needs to be modified so that it is independent of the extrinsic parameters. So the match is defined as the inner product maximised over the time of coalescence and phase of the waveform and can be expressed as
\begin{equation}\label{match}
    \mathcal{M}(h_{1}, h_{2}) = \max_{\phi_{c}, t_{c}} (h_{1}|h_{2}),
\end{equation}
where the inner product is defined as
\begin{equation}
    (h_{1}|h_{2}) = 4 \Re \int_{0}^{\infty} \frac{\tilde{h}_{1}^{*}(f)\tilde{h}_{2}(f)}{S_{h}(\textit{f})} df,
\end{equation}

$S_{h}(\textit{f}$) is the one-sided noise power spectral density (PSD) while $h_{1}$ and $h_{2}$ are the two gravitational wave templates with unit norm \citep{Brown2012}. The noise observed in a gravitational wave detector is modelled as stationary which means that the frequencies are independent and therefore the one-sided noise power spectral density, $S_{n}(f)$, can be defined as  
\begin{equation}\label{average noise}
\langle \tilde{n}^{*}(f) \tilde{n}(f') \rangle = \delta(f-f')\frac{1}{2}{S_{n}(\textit{f})},
\end{equation}

where $\tilde{n}(f)$ is defined as the Fourier transformed detector noise in the frequency domain and $\tilde{n}*(f)$ is the complex conjugate. Equation \ref{match} can intuitively be described as how similar two gravitational-wave templates are. If the two templates have similar parameters (i.e. almost equivalent mass and spins) then it is a strong match, $\sim1$ because the templates are almost identical. Whereas, if the two templates have very different parameters then it is a weak match, $\sim0$. The mapping between the intrinsic parameters of a compact binary merger and the match value can be described a manifold, which in this research has been called the \textit{match manifold} \citep{BenOwen1996}. The \textit{match manifold} is a continuous manifold but it's sensitivity to different intrinsic parameters varies. In the low mass regions of binary black hole mergers, small changes in the mass result in significant changes in the calculated match value. In contrast, in the high mass regions of binary black hole mergers, small changes in the mass result in insignificant changes in the calculated match value. This behaviour of the match manifold can be depicted in Figure \ref{fig: Match Manifold} and is the reason why template banks contain lots of templates in the low mass region compared to the high mass regions

%% file: LearningMatch.tex
\begin{figure}
    \centering
    \includegraphics[scale=0.4]{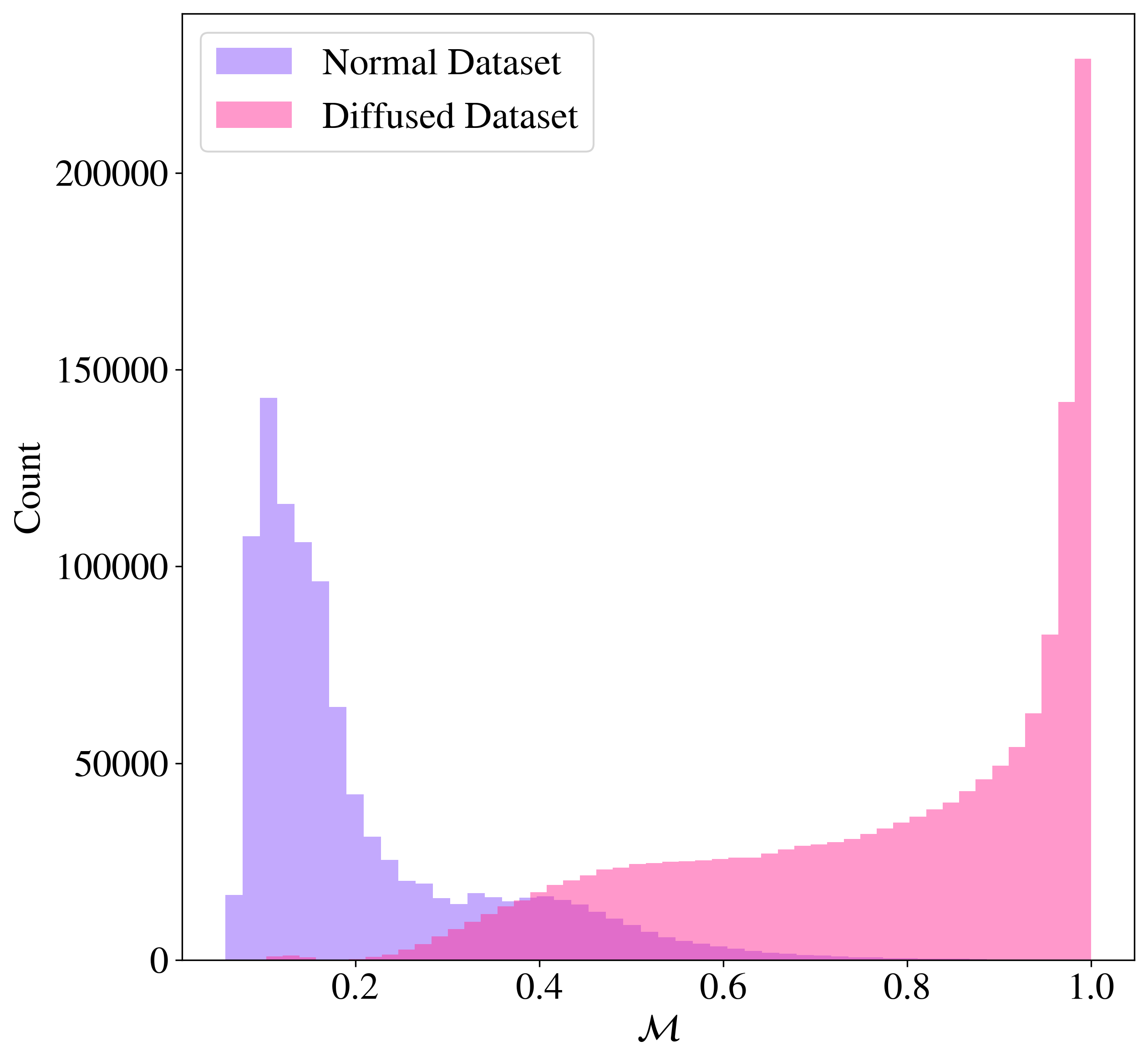}
    \caption{A histogram of matches in the training dataset. The diffused dataset resulted in higher match values compared to the normal dataset.}
    \label{fig: Dataset equal aligned spin}
\end{figure}

\begin{figure}
    \centering
    \includegraphics[scale=0.4]{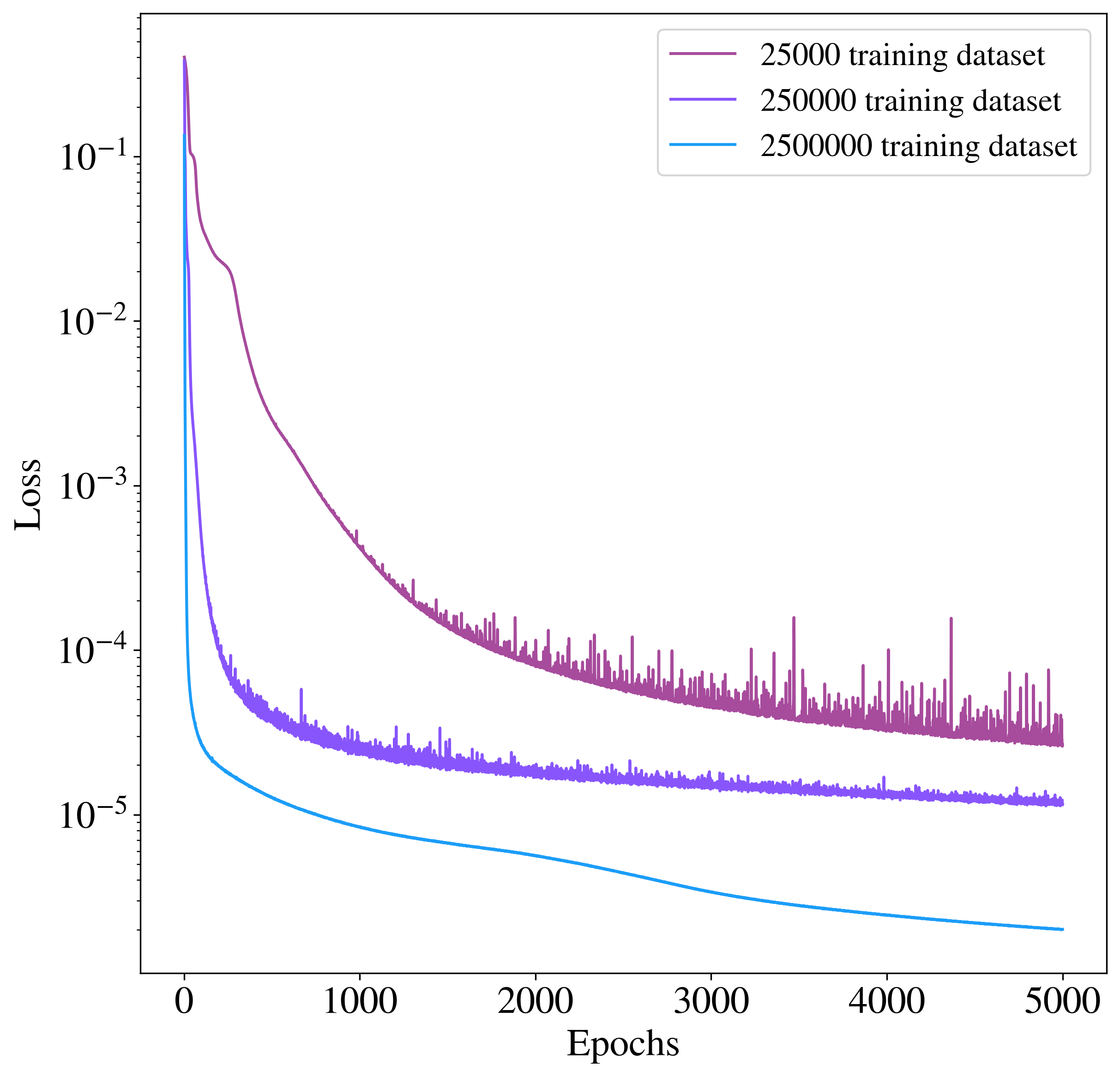}
    \caption{The training loss curves of the same LearningMatch model with three different sized training datasets.}
    \label{fig: Dataset Loss Curve}
\end{figure}

\begin{figure*}
    \centering
    \includegraphics[scale=0.55]{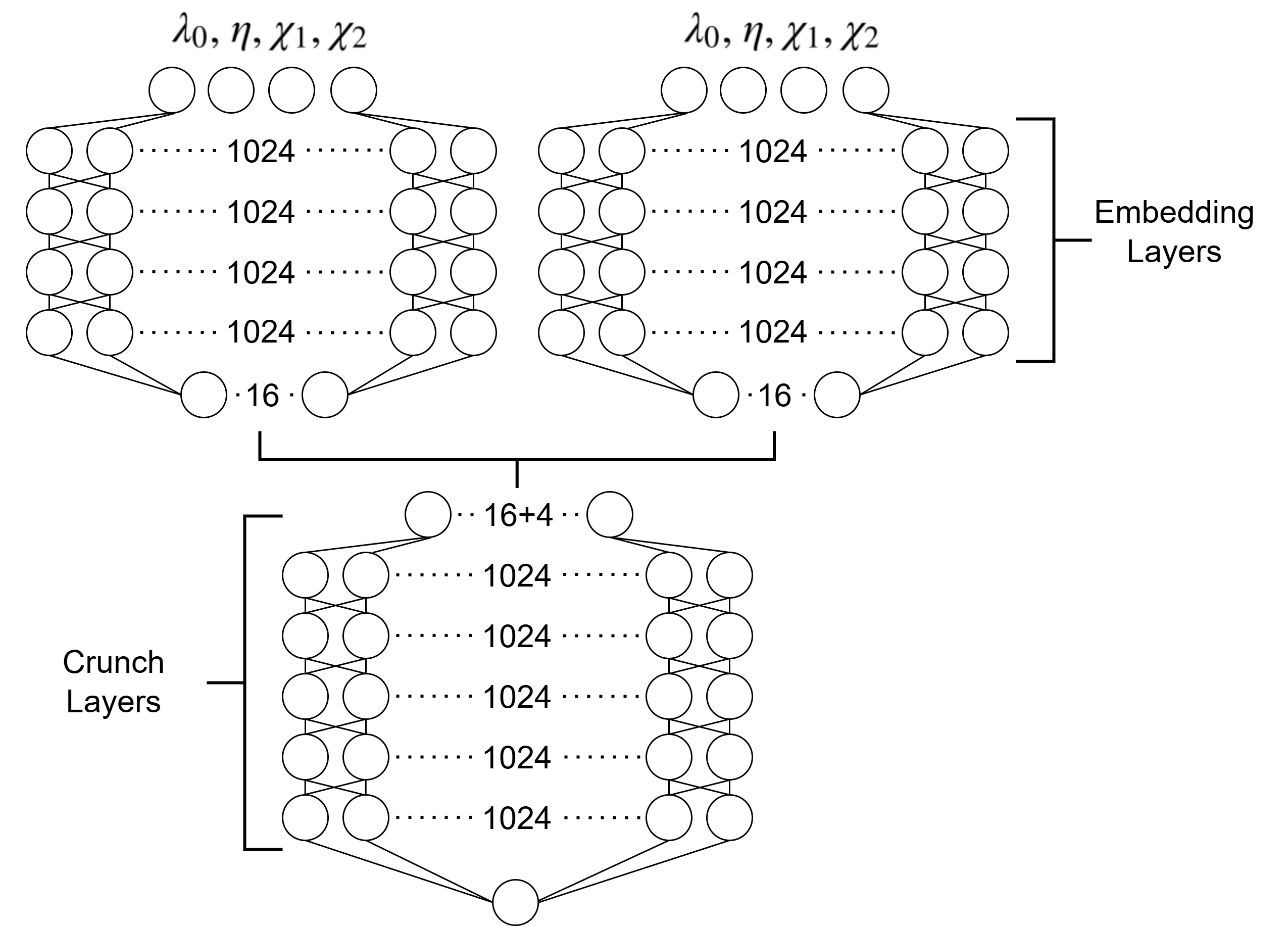}
    \caption{A schematic diagram of LearningMatch's architecture. The embedding layers are used to embed the parameters of a gravitational-wave template into a space that is easier to learn. The output of the embedding layers is then subtracted from each other and squared. This is then combined with the average of the original parameters of the gravitational-wave template (4 additional dimensions) to become the input for the crunch layers. The output of the crunch layer is the match. A Rectified Linear Unit activation function was used between each layer and has been omitted for clarity \citep{agarap2018deep}.}
    \label{fig: LearningMatch Architecture}
\end{figure*}

\begin{figure}
    \centering
    \includegraphics[scale=0.39]{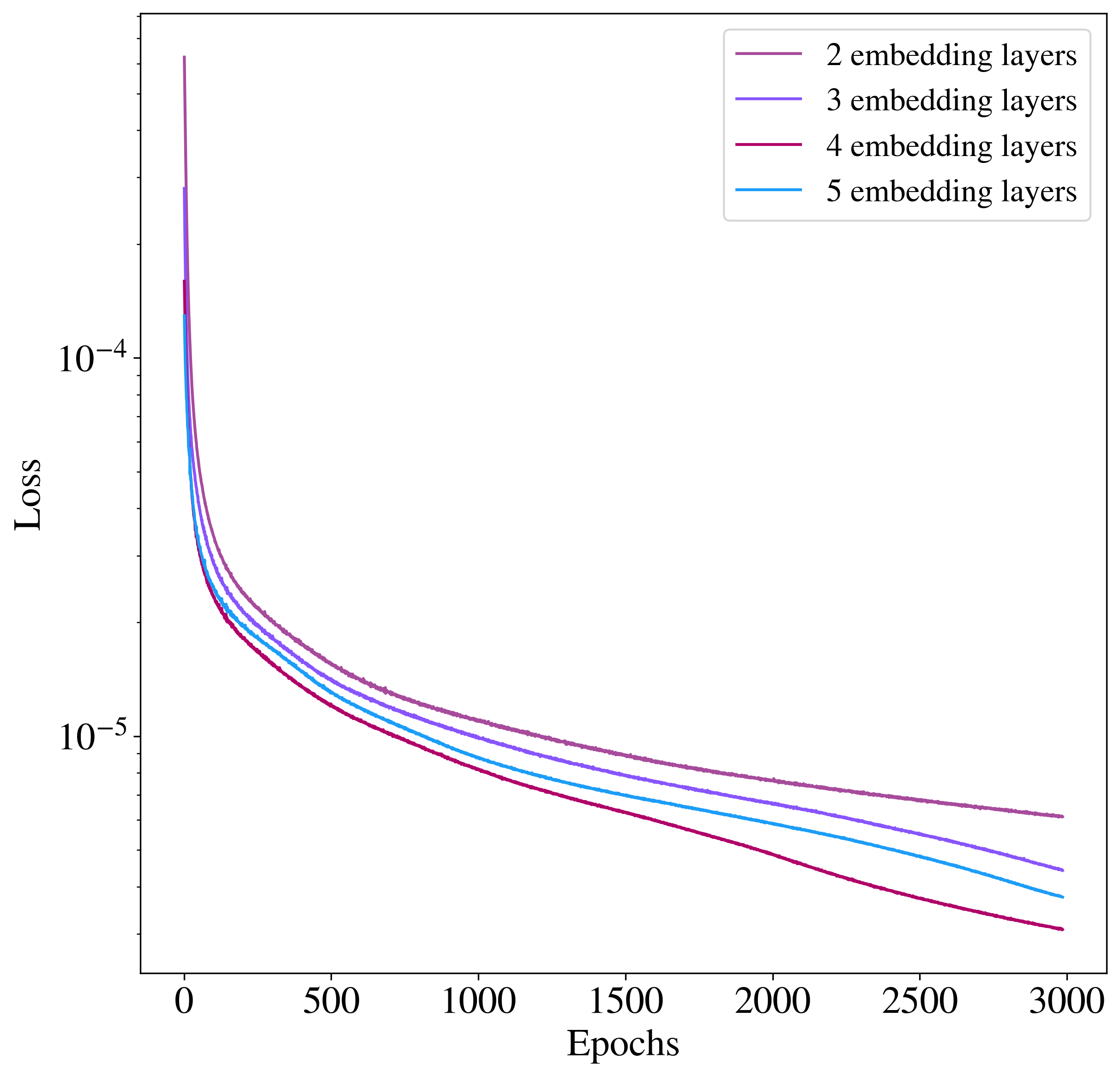}
    \caption{The training loss of the same LearningMatch model with different depths of the embedding layer.}
    \label{fig: Embedding Loss Curve}
\end{figure}

\begin{figure}
    \centering
    \includegraphics[scale=0.4]{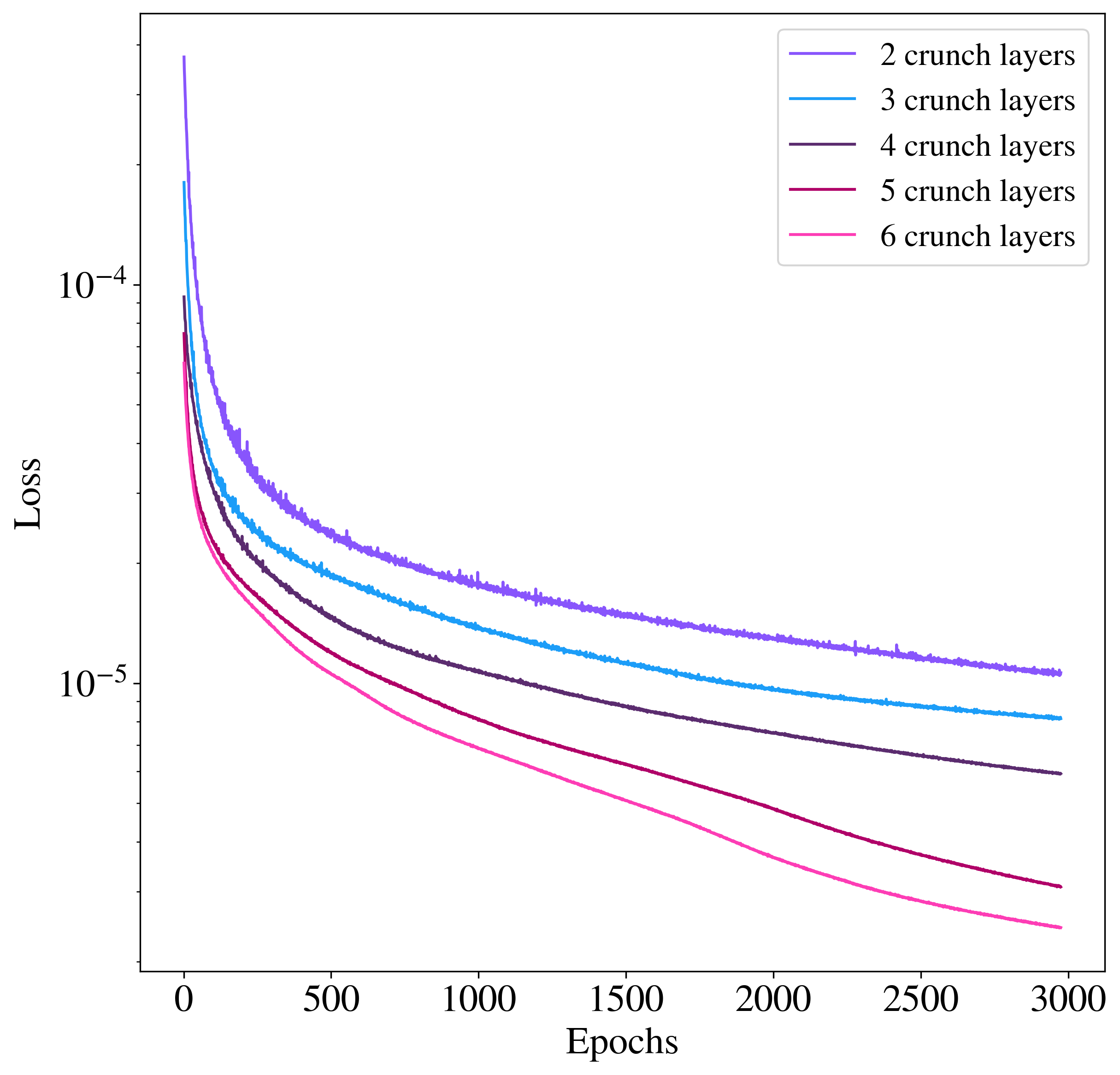}
    \caption{The training loss curves of the same LearningMatch model with varying depths of the crunch layer.}
    \label{fig: Crunch Loss Curve}
\end{figure}

\begin{figure}
    \centering
    \includegraphics[scale=0.4]{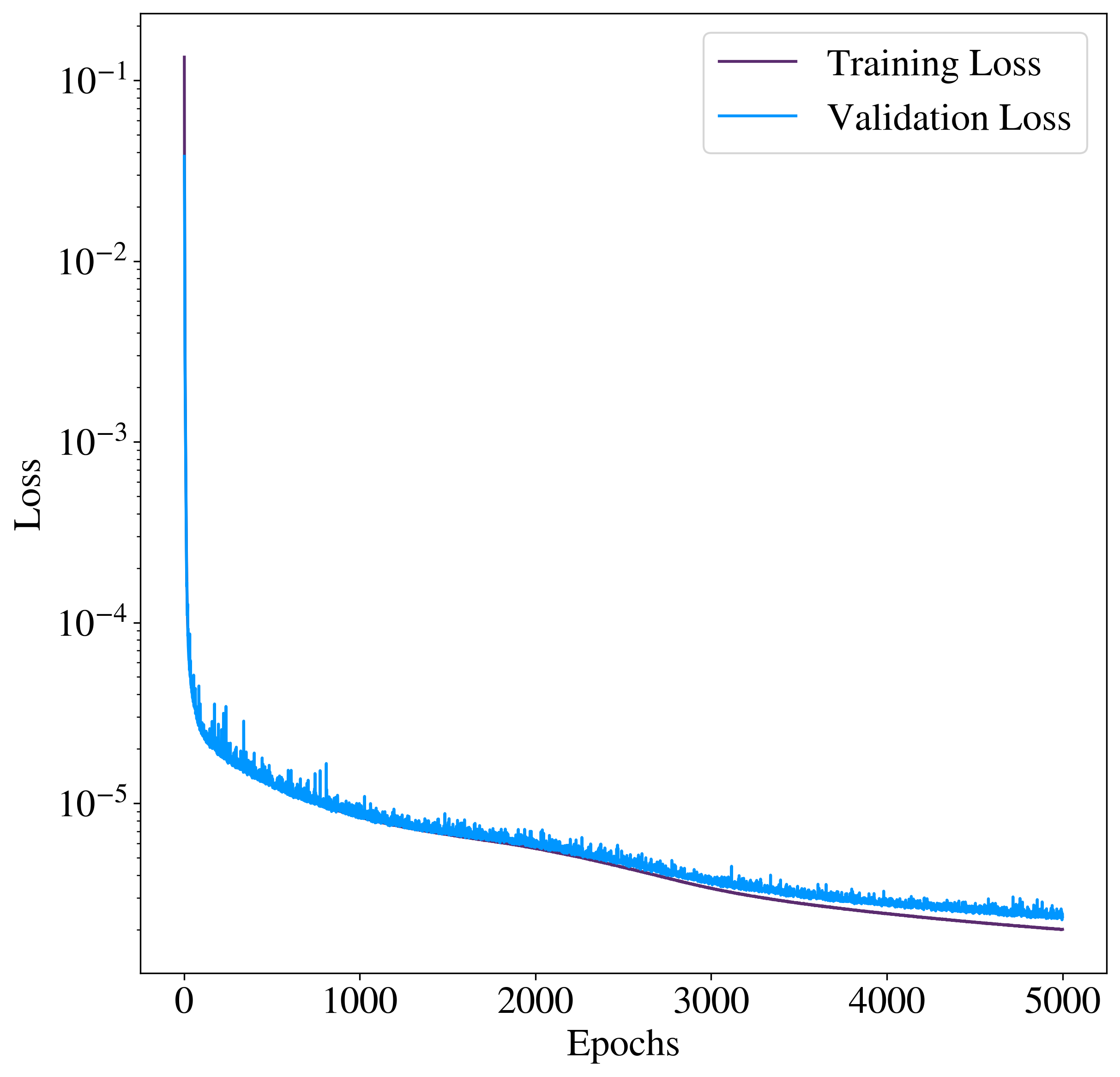}
    \caption{The training and validation loss curves for the LearningModel, described in this paper, that was trained on the equal aligned spin dataset.}
    \label{fig: Equal Aligned Spin Loss Curve}
\end{figure}

\subsection{The Dataset}
LearningMatch is a Siamese neural network that has learned the mapping between the parameters of two binary black hole templates and the match, therefore replacing Equation \ref{match} (see Section \ref{Theory} for more information). The dataset was created using PyCBC, which generates the two templates with different parameters and then calculates the match between them using the standard mathematical procedure outlined in Section \ref{Theory} \citep{PyCBC}. A simulated PSD representative of the third observing run was used and accessed using PyCBC (called \textit{aLIGO140MpcT1800545} in the software package) \citep{PyCBCPSD}. A sample frequency of 2048 Hz, signal duration of 32 s, and a low frequency cut-off of 15 Hz was used to generate the PSD. This PSD was reasonably representative of the detector when this paper was written. The template family used in this research was `IMRPhenomXAS' \citep{IMRPhenomXAS} and the match was generated with a low frequency cut-off of 18 Hz. The template parameters used to train LearningMatch were  $\lambda_{0}$ (which is proportional to the chirp mass), $\eta$ (symmetric mass ratio), and equal aligned spin ($\chi_{1}$ = $\chi_{2}$). $\lambda_{0}$ is defined as

\begin{equation}
\lambda_{0} = \left( \frac{\mathcal{M}_{c}}{\mathcal{M}_{c_{ref}}} \right)^{-5/3}, 
\end{equation}

where $\mathcal{M}$ is the chirp mass and $\mathcal{M}_{ref}$ is reference chirp mass. $\mathcal{M}_{c}$ is called the chirp mass and can be expressed as 

\begin{equation}
\mathcal{M}_{c} = \frac{(m_{1}m_{2})^{3/5}}{(m_{1}+m_{2})^{1/5}}, 
\label{chirp mass}
\end{equation}

where $m_{1}$ and $m_{2}$ are the mass of the two black holes. $\mathcal{M}_{ref}$ was predefined to be 5 $M_{\odot}$ so that $\lambda_{0}$ ranged between 0-1. This range is preferred when training neural networks because it results in the gradients being constrained in backpropagation, an algorithm used to train neural networks \citep{Brownleee2020, Bishop1996}. $\mathcal{M}_{c}$ varied between $5 M_{\odot}$ and $20 M_{\odot}$ which corresponds to $\lambda_{0}$ varying between 0.006 and 1.4. $\eta$ is called the symmetric mass ratio which can be expressed as $\eta = \frac{m_{1}m_{2}}{(m_{1}+m_{2})^{2}}$ and varied between 0.1 and 0.24999. In this mass range, the Siamese neural networks learned $\lambda_{0}$ and $\eta$ the best compared to other mass parameters because $\lambda_{0}$ smooths out the low chirp mass regions. As described in Section \ref{Theory}, the \textit{match manifold} changes significantly in the low mass regions and therefore the Siamese neural network is required to learn these details well. However, research has shown that neural networks in general struggle in learning details in data and therefore $\lambda_{0}$ was used to enable the Siamese neural network to learn the low chirp mass region well \citep{SpectralBias, machinelearningbias}. Equal aligned spin that varied between -0.99 and 0.99 was used in this research because the accuracy desired could not be achieved using random aligned spin ($\chi_{1} \neq \chi_{2}$). This will be discussed further in Section \ref{Discussion}.

Two methods were used to generate the datasets required to train LearningMatch. The first method chose $\lambda_{0}$, $\eta$ and equal aligned spin parameters from a uniform distribution for both templates and then the match was calculated between these templates. The second method generated a template with $\lambda_{0}$, $\eta$ and equal aligned spin chosen from a uniform distribution. Then the other template was generated using $\lambda_{0}$, $\eta$ and equal aligned spin at different standard deviations. The standard deviations used for the second template were $\lambda_{0} = 0.0001$, $\eta = 0.01$, and $\chi_{1}=\chi_{2}=0.01$ and $\lambda_{0} = 0.001$, $\eta = 0.01$, and $\chi_{1}=\chi_{2}=0.01$ and $\lambda_{0} = 0.001$, $\eta = 0.02$, and $\chi_{1}=\chi_{2}=0.01$. These standard deviations were determined using empirical tuning to identify which combination produced the largest number of match values close to 1. This dataset was called the \textit{diffused dataset} because for each template that had parameters chosen from a uniform distribution, three templates were chosen nearby with different standard deviation combinations. The \textit{diffused dataset} resulted in a dataset that focused on the match values near one, see Figure \ref{fig: Dataset equal aligned spin}. Both methods were used to generate the training, validation and test dataset. The first method was used to create datasets of the order of 1 million (training dataset), 100,000 (validation dataset), 100,000 (test dataset) in size. The second method was used to create datasets of the order of 1.5 million (training dataset), 150,000 (validation dataset), 150,000 (test dataset) in size. These two methods were then combined to create the training, validation and test datasets that consist of 2.5 million, 250,000, and 250,000 data points, respectively. It was concluded that 2.5 million data points were required to train the LearningMatch to the desired accuracy, see Figure \ref{fig: Dataset Loss Curve}. The datasets for this research can be accessed on Zenodo \citep{susanna_green_zenodo}.

\subsection{The Architecture}
Siamese neural networks consist of two subnetworks that that get combined in a final output neural network. In LearningMatch the subnetworks were called the \textit{Embedding Layers} because they were used to transform the template parameters, see Figure \ref{fig: LearningMatch Architecture} for more detail. Whilst the final output of the neural network was called the \textit{Crunch Layers} because this would be used to determine the match. $\lambda_{0}$, $\eta$ and equal aligned spin (i.e. $\chi_{1}$=$\chi_{2}$) of both templates are used as the input to the \textit{Embedding Layers}. The output of the \textit{Embedding Layers} is then combined with the difference between template parameters squared and the average of all the template parameters. The output of the \textit{Embedding Layers}, the difference squared, and the average is used as the input for the \textit{Crunch Layers}. The reason for using the difference squared is to emphasise the fact that when the template parameters are similar, the match is close to 1. The average is used to enable the neural network to learn the relative position of the templates on the match manifold. The \textit{Embedding Layers} consist of 4 layers of 1024 neurons, and the \textit{Crunch Layers} consist of 5 layers of 1024 neurons. Rectified Linear Unit (ReLU) activation functions are used between each layer of the \textit{Embedding Layers} and \textit{Crunch Layers} \citep{ReLU1941, agarap2018deep}. ReLU was used because it performed the best compared to other activation functions (activation functions help neural networks to learn by adding non-linearity into the training process). A Linear layer is used for the final layer in both \textit{Embedding Layers} and \textit{Crunch Layers}. The architecture of LearningMatch was determined by trial and error. We found that the depth of the \textit{Embedding Layers} did not have a significant impact on the accuracy of LearningMatch, see Figure \ref{fig: Embedding Loss Curve}. We also found that 5 layers in the \textit{Crunch Layers} was the best depth because LearningMatch's performance did not improve significantly with 6 layers, see Figure \ref{fig: Crunch Loss Curve}. Additional layers in the LearningMatch model results in an increased time to predict the match. Various experiments were conducted using different neural network architectures and the one presented in this paper was determined to be the fastest and the most accurate.

\subsection{The Training}
LearningMatch required 5000 epochs to train which took approximately 31 hours on a A100 GPU using PyTorch \citep{NEURIPS2019_9015}. The mean squared error was chosen as most appropriate loss function because predicting the match can be viewed as a simple regression problem, where the actual match calculated using Equation \ref{match}, $\mathcal{M}_{actual}$, is compared with the match predicted by the Siamese neural network, $\mathcal{M}_{predicted}$. The mean squared error was used to determine the accuracy of this prediction and is defined as
\begin{equation}
MSE = \frac{1}{n} \sum^{n}_{i=1} (\mathcal{M}_{actual}-\mathcal{M}_{predicted})^{2},
\end{equation}
where \textit{n} is the number of data points. An Adam optimiser with an initial learning rate of $10^{-6}$ was then used to minimise the loss function and update the parameters in the Siamese neural network \citep{Adam2014}. A batch size of 1024 was used on the training and validation datasets due to the size of the datasets. During the training process, it was noted that LearningMatch never overtrained given the large number of epochs. 5000 epochs were sufficient to train LearningMatch without the model becoming overtrained. Overtraining would result in the Siamese neural network memorising the data and therefore would not be able to predict the match well on unseen data, otherwise known as generalisation. As shown in Figure \ref{fig:  Equal Aligned Spin Loss Curve}, the training and the validation loss curves do not significantly separate which is a typical characteristic of an overtrained model.

%% file: Results.tex
\begin{figure}
    \centering
    \includegraphics[scale=0.4]{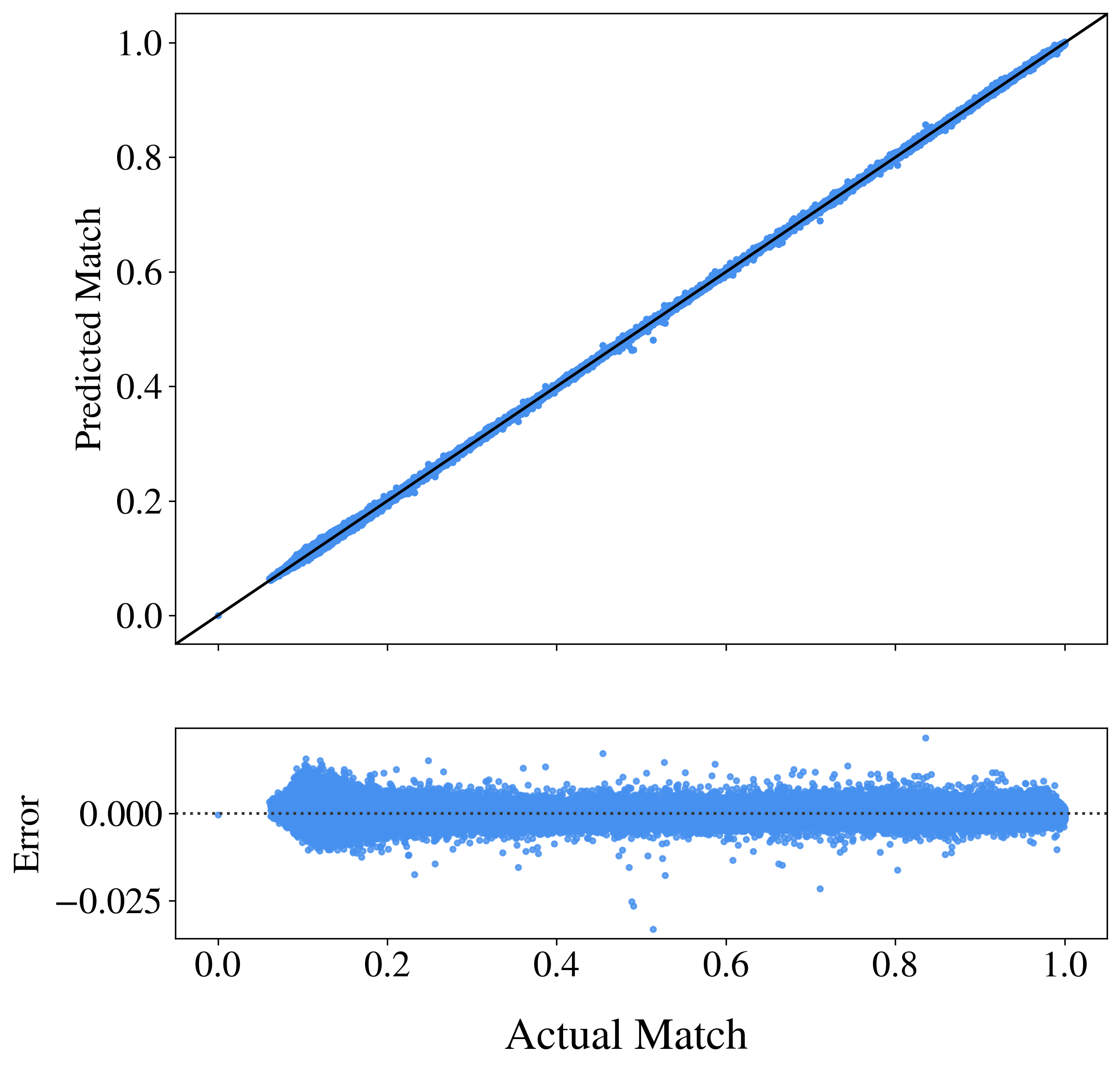}
    \caption{The comparison between the actual match, which was calculated using the mathematical procedure outlined in section \ref{Theory}, and the predicted match, which is the match determined by LearningMatch, using an unseen test dataset. The error is defined as the difference between the actual match and the predicted match.}
    \label{fig: Mass and Equal Spin Actual Predicted}
\end{figure}

\begin{figure}
    \centering
    \includegraphics[scale=0.4]{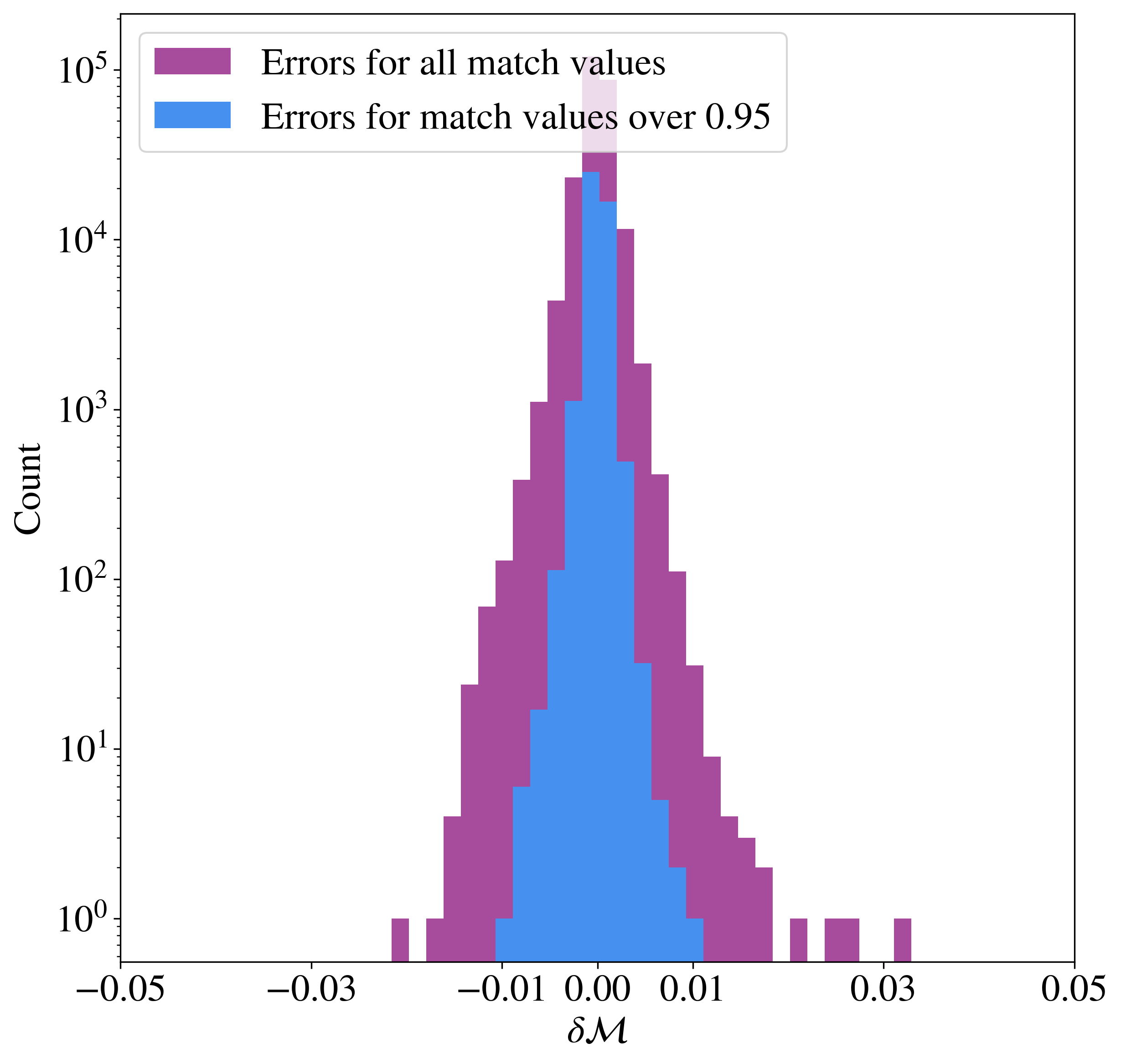}
    \caption{The error distribution for LearningMatch on a unseen test dataset. The error is defined as the actual match (calculated using mathematical methods) subtracted from the match predicted by the Siamese neural network. Due to the potential application of LearningMatch, the actual match values greater than 0.95 were highlighted in this plot to showcase the accuracy in this region.}
    \label{fig: Mass and Equal Spin Error Distribution}
\end{figure}

\begin{figure}
    \centering
    \includegraphics[scale=0.4]{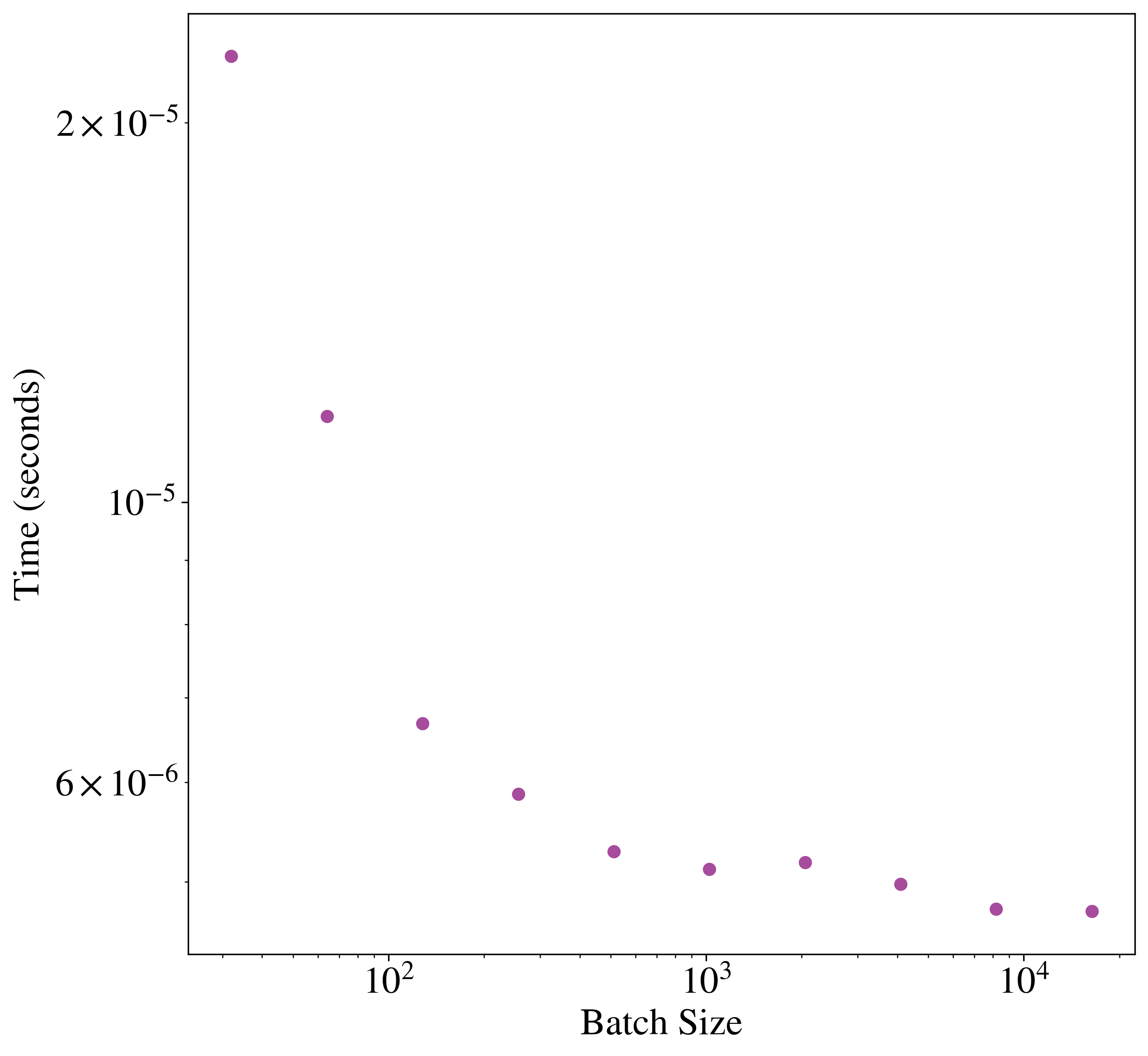}
    \caption{The time taken for LearningMatch to predict the match depending on the batch size. The larger the batch size, the more times the match is predicted. However, the time taken to predict a single match is reduced.}
    \label{fig: Time Taken to predict Match}
\end{figure}

A trained LearningMatch can predict the match to within $3.3\%$ of the actual match value, determined by standard mathematical calculations defined in Section \ref{Theory}. As shown in Figure \ref{fig: Mass and Equal Spin Actual Predicted}, LearningMatch has learned this accuracy for the entire match manifold. For actual match values greater than 0.95, a trained LearningMatch model can predict the match to within $1\%$ of the actual match value, as summarised in Figure \ref{fig: Mass and Equal Spin Error Distribution}. The accuracy in this region (match values greater than 0.95) is important because of the potential applications of LearningMatch which are discussed in Section \ref{Discussion}. Using PyCBC, an open source gravitational-wave software, the match can be calculated in 40 ms on a single CPU core. PyCBC uses the standard mathematical procedures described in Section \ref{Theory} to calculate the match, which involve templates being generated and then the match being calculated. LearningMatch can predict the match in 20 $\mu s$ (mean maximum value) on a single A100 GPU, see Figure \ref{fig: Time Taken to predict Match}. Therefore, LearningMatch is at least 3 orders of magnitude faster at predicting the match than standard mathematical procedures.

%% file: Discussion.tex
LearningMatch is a Siamese neural network that can predict the match to within $3.3\%$ of the actual match value, as seen in Figure \ref{fig: Mass and Equal Spin Actual Predicted}. For match values greater than 0.95, LearningMatch can predict the match to within $1\%$ of the actual match value which is shown in Figure \ref{fig: Mass and Equal Spin Error Distribution}. The match can be predicted by LearningMatch in 20 $\mu$ (mean maximum value) on a single A100 GPU. Whereas, PyCBC can calculate the match using standard mathematical procedures (which involve generating two gravitational-wave templates and then calculating the match between them) in 40 ms on a single CPU core. Therefore, LearningMatch is at least 3 orders of magnitude faster at determining the match than current calculations however, this number does exclude the ~2 days required to train LearningMatch on a single A100 GPU. LearningMatch is not a computationally feasible option when only one or two calculations of the match are needed. However, LearningMatch is computationally feasible when lots of match calculations are required because the time taken for a single match calculation is reduced with larger number of match calculations, as shown in Figure \ref{fig: Time Taken to predict Match}. Speed and accuracy were the main objectives of this research so that LearningMatch could be integrated into gravitational-wave data-analysis pipelines that require a lot of match calculations, such as template bank generation.

The match is can be computationally expensive when the gravitational-wave templates are computationally expensive to generate. In this research, a Siamese neural network is used to surpass template generation and predict the match. Template banks are used to detect the gravitational-wave signal from the interferometric strain data using matched-filtering techniques \citep{Allen2021OptimalBanks, Adams2016, PyCBCSearch2016, Kovalam2022EarlySearch, Sakon2024TemplateKAGRA, GstLAL_templatebank2021}. There are various ways of generating template bank, with geometric and stochastic algorithms being most favoured by the literature \citep{Roulet2019TemplateAlgorithm, Cokelaer2007GravitationalSignals, Prix2007Template-basedSpaces, Harry2009, Allen2022PerformanceBanks, Babak2006}. In the stochastic algorithm, a template with new parameters is proposed and the match is computed with all the templates already in the template bank; if the match is greater than 0.97 the template is accepted into the bank. LearningMatch could be integrated into this algorithm to speed up the match computation and therefore the time required to generate a template bank could be significantly reduced. LearningMatch learned a specific black hole mass range and equal aligned spin therefore this LearningMatch model could only generate template banks that could identify binary black hole mergers \citep{Roy2017templatebank}. This work could be extended to explore how well LearningMatch learns other mass ranges, such as the binary neutron star mass range. 

During our research, the mass range had to be constrained because LearningMatch failed to learn the higher mass regions. This is because $\lambda_{0}$ is a parameter that results in the templates being more uniformly spaced in mass and when the training datasets are generated using a uniform distribution in $\lambda_{0}$, the high-mass regions contained too few data points. We tried generating the training datasets using a $m_{1}$ and $m_{2}$ mass parameterisation but, the lower mass regions then contained too few data points. This is a limitation of LearningMatch because different parameters will be required to learn different areas in the mass regions. Otherwise, more research will be required to identify the correct parametrisation of the template parameters. Hypothetically, LearningMatch could also be used to learn more complicated match manifolds, such as precessing and eccentric compact binary mergers but this will require thorough scrutiny to make sure the entire manifold has been learned \citep{Brown2012NonspinningApproximation, Harry2014, Ajith2014EffectualSpins, Brown2010EffectDetectors, Csizmadia2012GravitationalBinaries, Coughlin2015DetectabilityDetectors, KhunSangPhukon2025}.

This work could be extended so that LearningMatch learns the mapping between $\lambda_{0}$, $\eta$, and random aligned spin ($\chi_{1}$ $\neq$ $\chi_{2}$) and the match. Initially in this research we tried to get LearningMatch to learn this match manifold but the Siamese neural network did not learn the entire match manifold. Some areas of the match manifold were learned to a lesser degree of accuracy than desired (or not at all). If future work modified the Siamese neural network presented in this paper so that the accuracy could be achieved, this LearningMatch model could be integrated into the Markov Chain Monte Carlo methods and therefore speed up gravitational-wave parameter estimation pipelines \citep{Meyer2001UsingData, vanderSluys2008Gravitational-WaveBinaries, VanDerSluys2008ParameterCarlo, Raymond2010TheSignals}. Markov Chain Monte Carlo methods are used to recover the parameters of the gravitational wave once the gravitational-wave signal has been identified \citep{Meyer2001UsingData, vanderSluys2008Gravitational-WaveBinaries, VanDerSluys2008ParameterCarlo, Raymond2010TheSignals}. LearningMatch could be used to predict the match of two gravitational-wave templates in the Markov Chain Monte Carlo method which would make proposed steps more efficient. LearningMatch integrated into Markov Chain Monte Carlo methods could be particularly useful for third generation detectors where template generation is computationally expensive for certain gravitational-wave sources \citep{Lenon2021EccentricExplorer}.

%% file: Conclusions.tex
LearningMatch is a Siamese neural network that has learned the match manifold. The match is defined as how similar two gravitational-wave templates are with different parameters and can be calculated using the standard definition outlined in Section \ref{Theory}. LearningMatch learned the mapping between $\lambda_{0}$, $\eta$ and equal aligned spin ($\chi_{1}$ = $\chi_{2}$) of two gravitational-wave templates and the match. We showed that LearningMatch can predict the match to within $3.3\%$ of the actual match value (i.e. the match value calculated using standard definitions). For match values greater than 0.95, LearningMatch can predict the match to within $1\%$ of the actual match value. We also showed that LearningMatch is 3 orders of magnitude faster at predicting the match between two gravitational-wave templates compared current calculations used in the gravitational-wave community. LearningMatch was designed to be integrated into gravitational-wave data-analysis pipelines that require lots of match calculations, such as template bank generation, so speed and accuracy were the main objectives in this research.

Template banks are used to identify potential gravitational-wave signals in the interferometric strain data by matched-filtering. Template banks for binary black hole mergers are generated using equal aligned spin to reduce the computational cost of matched-filtering. An equal aligned spin LearningMatch model could be integrated into a stochastic template bank generation algorithm and therefore a template bank could be generated quickly. This research could be extended to learn the mapping between the $\lambda_{0}$, $\eta$ and random aligned spin ($\chi_{1} \neq \chi_{2}$). We initially tried to get a Siamese neural network to learn this match manifold but the desired accuracy of approx. $1\%$ could not be achieved. A LearningMatch model trained on random aligned spins could be integrated into a Markov Chain Monte Carlo method as a fast algorithm to reject suggested proposals (or conversely suggest new proposals). This would result in more efficient sampling of gravitational-wave data and therefore reduce the time taken to recover the gravitational-wave parameters. In theory, LearningMatch has the potential to learn the match manifold of other gravitational-wave sources, such as binary neutron star mergers, and more complicated match manifolds, such as precessing compact binary mergers. This would mean that our understanding of gravitational-wave sources could be extended. LearningMatch could also be integrated into the data-analysis pipelines for future gravitational-wave detectors. In this paper we explored the feasibility of replacing a mathematical equation with a neural network and this idea could also be applied to other areas gravitational-wave data-analysis pipelines.